\documentclass[runningheads]{llncs}
\usepackage{graphicx}
\usepackage[inline]{enumitem}
\usepackage{stmaryrd} 
\usepackage{url} 
\usepackage{color,soul}
\usepackage{booktabs} 

\begin{document}
\title{MS-Shift: An Analysis of MS MARCO Distribution Shifts on Neural Retrieval}
%
\titlerunning{An Analysis of MS MARCO Distribution Shifts}
%



\author{Simon Lupart \inst{1} \and
Thibault Formal \inst{1,2} \and Stéphane Clinchant \inst{1} }
\authorrunning{S. Lupart et al.}
\institute{Naver Labs Europe, Meylan, France \\
\email{\{firstname.lastname\}@naverlabs.com} \and Sorbonne Université, ISIR, Paris, France}
\maketitle              
\begin{abstract}
Pre-trained Language Models have recently emerged in Information Retrieval as providing the backbone of a new generation of neural systems that outperform traditional methods on a variety of tasks.
However, it is still unclear to what extent such approaches generalize in zero-shot conditions. The recent BEIR benchmark provides partial answers to this question by comparing models on datasets and tasks that differ from the training conditions. We aim to address the same question by comparing models under more \emph{explicit} distribution shifts. To this end, we build three query-based distribution shifts within MS MARCO (query-semantic, query-intent, query-length), which are used to evaluate the three main families of neural retrievers based on BERT: sparse, dense, and late-interaction -- as well as a monoBERT re-ranker.
We further analyse the performance drops between the train and test query distributions. In particular, we experiment with two generalization indicators: the first one based on train/test query vocabulary overlap, and the second based on representations of a trained bi-encoder. Intuitively, those indicators verify that the further away the test set is from the train one, the worse the drop in performance. We also show that models respond differently to the shifts -- dense approaches being the most impacted.
Overall, our study demonstrates that it is possible to design more \emph{controllable} distribution shifts as a tool to better understand generalization of IR models. Finally, we release the MS MARCO query subsets, which provide an additional resource to benchmark zero-shot transfer in Information Retrieval.

\end{abstract}
%
%
%
\section{Introduction}

The ability of machine learning models to generalize to unseen cases under distribution shifts remains a major challenge and concern for systems deployed in the real world~\cite{csurka2021unsupervised}. 
In Information Retrieval (IR), the question of generalization has often been eluded, due to the robust and long-standing performance of term-based approaches~\cite{10.1145/3331184.3331340}. However, with the recent advent of neural IR based on Pre-trained Language Models (PLM) like BERT~\cite{bert}, the generalization issue has become as relevant as ever, as recently shown in the zero-shot BEIR benchmark~\cite{beir_2021}. By comparing various types of BERT-based models on different domains and tasks, Thakur et al. show how cross-encoders, as well as retrieval models with lexical priors such as doc2queryT5~\cite{doct5} or ColBERT~\cite{colbert}, tend to be more robust, while dense bi-encoders such as DPR~\cite{karpukhin2020dense} or TAS-B~\cite{Hofstaetter2021_tasb_dense_retrieval} seem to suffer more from domain shifts -- with performance lower than BM25 overall. Knowing that many production systems use (or will use) models based on PLM (e.g.~\cite{xiong2021approximate}), while being exposed to new documents and queries every day, robustness is thus a critical aspect that must be assessed.

Outside of the IR field, Wiles et al.~\cite{finegrained_ds_neurips21} propose a framework to evaluate computer vision models under various distribution shifts, in order to assess the important aspects for which robustness is required, and which models are effectively robust. Intuitively, a dataset is composed of samples with various attributes (for instance, color, shape, or lightning), where some attribute values would be seen at training time -- and others not. The objective is then to be able to learn representations that are invariant to such variations, to better transfer to unseen attributes. Our objective echoes the same research question, from the IR point of view, where \emph{terms} constitute the unit of variation -- their frequency in a training dataset having a potential impact on model effectiveness. 
Even if the BEIR benchmark implicitly defines various distribution shifts, we would like to \textit{explicitly} control the shifts to understand which of those are critical to evaluate robustness. In particular, we show that within MS MARCO, we can construct several distribution shift experiments that we believe will ease the study of these phenomena. Our main contributions are as follows:

\begin{itemize}
    \item we design and release multiple distribution shifts based on MS MARCO query attributes to help analysing the robustness to unseen attributes\footnote{Splits of MS MARCO queries available at \url{https://github.com/naver/ms-marco-shift}}; 
    \item we compare the three main families of first-stage retrievers based on BERT (dense or sparse bi-encoders, and late interaction), as well as a cross-encoder in those controlled shifts, and show that dense models are the most impacted;
    \item we analyse how the drops in effectiveness can be linked to the ``distance'' between train and test query distributions, in particular with two possible generalization indicators (term- and model-based). 
\end{itemize}

The structure of the paper is organized as follows: Section~\ref{sec:related} outlines prior works on generalization in IR. Section~\ref{sec: metho} details our methodology, while Section~\ref{sec:exp} contains the experimental setting. Results and analyses are reported in Section~\ref{sec:res}. Section~\ref{sec:conclu} summarizes the main conclusions and future research directions.

\section{Related Work}\label{sec:related}


Pre-trained Language Models (PLM) have impacted IR at its very core, owing to their ability to model complex semantic relevance signals, which makes them appealing to replace traditional term-based approaches in modern search engines. From re-rankers like monoBERT~\cite{passage_ranking} to models that directly tackle first-stage ranking -- including dense~\cite{karpukhin2020dense,xiong2021approximate,Hofstaetter2021_tasb_dense_retrieval,lin-etal-2021-batch,ren-etal-2021-rocketqav2} and sparse~\cite{10.1145/3404835.3463098,10.1145/3477495.3531857} bi-encoders, as well as late-interaction models~\cite{colbert,santhanam-etal-2022-colbertv2}, the effectiveness gains offered by such approaches is quite impressive. Initially evaluated on in-domain settings (like MS MARCO~\cite{msmarco}), where train and test queries follow the same distribution, conclusions became more contrasted when Thakur et al. released the zero-shot BEIR benchmark~\cite{beir_2021} -- in which some models like DPR~\cite{karpukhin2020dense} achieve lower overall performance than (unsupervised) BM25. In more detail, the BEIR benchmark consists of a test suite of 18 datasets -- each containing documents, queries, and corresponding \emph{qrels} -- that are used to evaluate models in zero-shot, i.e., without any sort of training based on those datasets. The selected datasets were chosen by three factors: diversity in tasks, domains, and difficulty. This makes BEIR really challenging as, differently from classical evaluation settings where the collection usually stays unchanged, here both queries and documents are ``new''. Furthermore, the task may also vary from the initial training objective (e.g., Question-Answering or Fact-Checking). By measuring the similarity between datasets/domains -- relying on the weighted Jaccard similarity~\cite{w_jaccard} between the document collections
-- the authors argued that BEIR indeed contains a diverse set of tasks, and is perfectly suited for zero-shot evaluation of neural IR models.  

In the meantime, other works have investigated various aspects of robustness or generalization capabilities of re-ranker models, in order to uncover their weaknesses or failure cases. First, there have been several works on the systematic testing of transformer-based rankers \cite{Cmara2020DiagnosingBW,Rau_howdiff,macavaney-etal-2022-abnirml,volske2021towards}. More specific to the IR field, the impact of shifting trends in search engines was studied on neural re-rankers as well (pre- and post-BERT), under the lens of catastrophic forgetting~\cite{catastrophic_ir_ecir21} and lifelong learning~\cite{https://doi.org/10.48550/arxiv.2201.03356}. Penha et al.~\cite{10.1007/978-3-030-99736-6-27} also analysed the robustness of various re-ranking models to typos (i.e. variations without changes of semantic) -- as search engines directly interact with users and may be exposed to such issues. Following works further complement the findings for dense bi-encoders against misspellings and paraphrasing~\cite{Zhuang2021DealingWT,DBLP:conf/sigir/SidiropoulosK22,10.1145/3477495.3531951,zhoualignement}. From a different perspective, various neural IR models (mostly re-rankers) have been shown to be vulnerable  under adversarial attacks -- usually by substituting words in documents or queries ~\cite{wu2022prada,wu2021neural,song-etal-2022-trattack,liu2022orderdisorder,Wang-Brittle}. Overall, such studies usually focus on a single type of model, and lack the comparison between the various architectures proposed to tackle \textit{first-stage} ranking, for which robustness might even be more critical.

A few works recently analysed the generalization (or zero-shot properties) of BERT-based first-stage rankers~\cite{rosa_modelsize}. Lexical matching has been shown to be architecture-dependent, especially in zero-shot~\cite{formal2021match}. Using the two train sets of MS MARCO and Natural Question, Ren et al.~\cite{https://doi.org/10.48550/arxiv.2204.12755} identify key factors that affect the zero-shot properties of dense models -- including the overlap between the source and target query sets, as well as the query type distribution. We similarly study the role of such overlap, but we rely on our \textit{explicit} shifts built within MS MARCO. Zhan et al.~\cite{https://doi.org/10.48550/arxiv.2204.11447} provide a thorough analysis on MS MARCO by \begin{enumerate*}[label=(\textit{\roman*})]
    \item identifying a strong train/test overlap within the dataset, that plagues the current evaluation of neural rankers,
    \item fixing this issue, relying on two new resampling strategies that allow to accurately compare zero-shot  properties of several retrieval architectures, on datasets without train/test overlaps,
    \item showing that bi-encoders fail to properly generalize compared to cross-encoders.
\end{enumerate*} While also being related, our work differs in that we aim to create \emph{controllable} subsets that would not only avoid train/test overlaps, but would also help to analyse the link between effectiveness drops and train/test similarity. Similar findings on leakages between Robust04 and MS MARCO have been recently outlined~\cite{https://doi.org/10.48550/arxiv.2206.14759}, which further motivates the need to build datasets with controlled shifts that do not contain such leakages. Other related works also investigated particularities of the MS MARCO collection, and possible bias the dataset could contain~\cite{Gupta_2022,https://doi.org/10.48550/arxiv.2112.03396,https://doi.org/10.48550/arxiv.2109.00062}. Finally, several methods have been proposed to adapt neural rankers to new domains or tasks -- usually without requiring supervised annotations on the targeted distributions~\cite{https://doi.org/10.48550/arxiv.2112.07577,https://doi.org/10.48550/arxiv.2209.11755,https://doi.org/10.48550/arxiv.2205.11498,zhan2022disentangled,coco-dr}. Such works which try to fix -- rather than understand the causes of -- generalization issues, are out of scope for our work, even though they can indirectly help on the understanding of generalization.

\section{Methodology}\label{sec: metho}

\subsection{Distribution Shifts}

\begin{table}[t]
\begin{center}
    \caption{Examples of queries from the topic clusters. We identified the following topics: Names and Public Figures, Dated Events, Pricing/Units, Medical Treatments and Biology/Physics.}
    \label{tab:query}
        \begin{tabular}{ll}
        \toprule
        \textbf{Topic} & \textbf{\quad \quad \quad \quad \quad \quad \quad \quad \quad \quad \quad  Queries} \\ \midrule
        $C_0$ & +what does the name brooke mean ; Camel Two Humps called ; \\
        & How Did George Peppard Die ; How Much is Bobby Brown Worth \\ \midrule
        $C_1$ & +when is mardi grai ; +which president has  
         living grandsons ; 23 is \\
         & what day of 2016 ; 2015 ncca footbal rankings \\ \midrule
        $C_2$ &  1 cm is how many millimeters ; . what is the major 
          difference between \\
          & a treaty and an executive agreement? ; 1 point perspective definition \\ \midrule
        $C_3$ & ECT is a treatment that is used for ; The ABO blood 
         types are examples \\
         & of ; The vitamin that prevents beriberi is ; 1.5 grams of sodium per day \\ \midrule
        $C_4$ & Ebolavirus is an enveloped virus, which means ; 
         \% of earths crust is \\
         & dysprosium ; +what is forbs as a food for animals? \\
        \bottomrule
    \end{tabular}
\end{center}
\end{table}


To investigate the behavior of models facing different types of shifts, we start from the MS MARCO passage dataset~\cite{msmarco}, which contains approximately $8.8M$ passages and $500k$ training queries. 
We then build upon it three shifts -- defined as a change of attributes distribution between train and test distribution -- from queries. Our goal is to cover all three \textit{lexical} (inferred from terms), \textit{semantic} (connected to meaning), and \textit{syntactic} shifts (related to word structure).
Additionally, note that shifts on training queries implicitly entail a shift in documents, as both relevant and negative-sampled documents used for training won't follow the same distribution.

\begin{figure}[t]
    \centering
	\includegraphics[width=0.6\columnwidth]{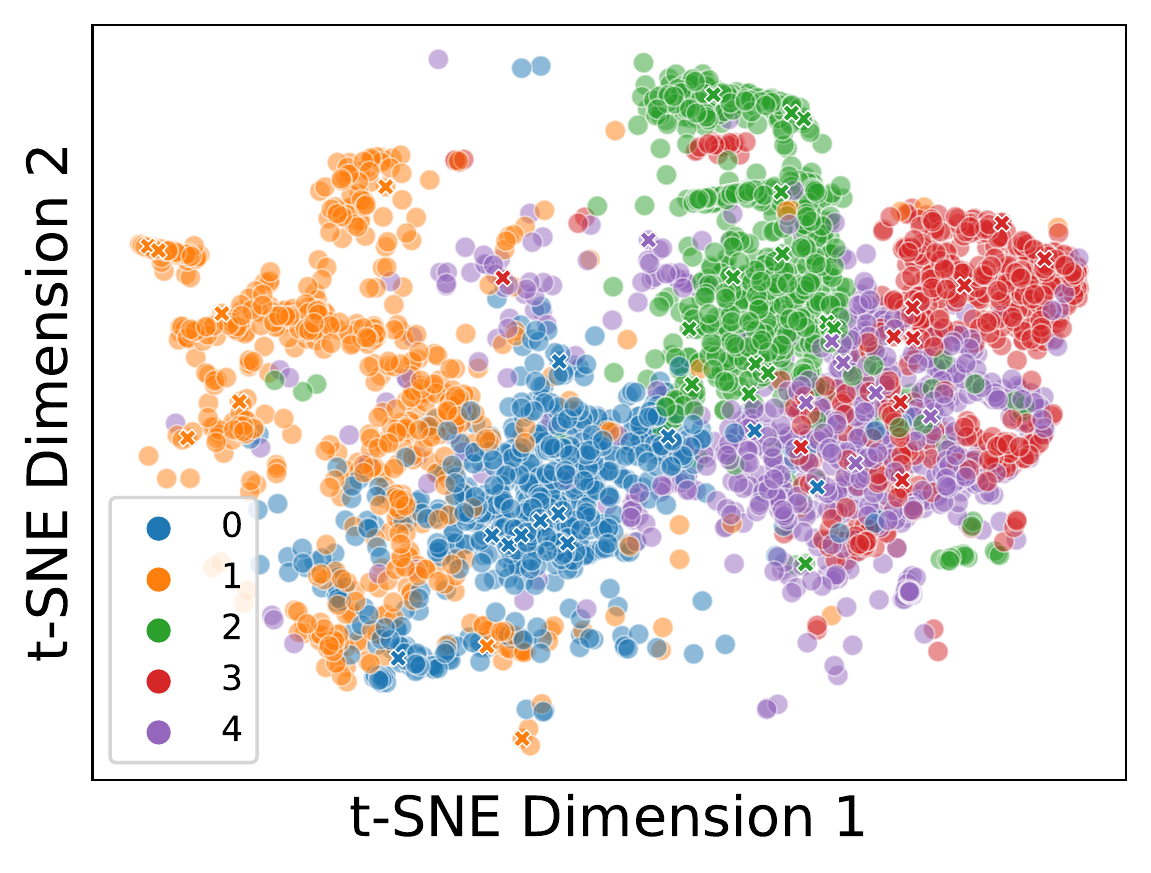}
	\caption{t-SNE on the query topics, where each cluster contains both training and evaluation queries. 
	Clusters 3 and 4 are close to each other but correspond to resp. Medical Treatments (red) and Biology/Physics (purple), the other cluster are well separated.}
	\label{fig:tsne}
\end{figure}

\paragraph{\bf{Query Topics}} We first propose to separate queries into five semantic clusters, referred to as $(C_i)_{i \in \llbracket 0,4\rrbracket}$, alongside their complements $\overline{C_i}=\{C_j | j \in \llbracket 0,4\rrbracket; j \ne i\}$. 
Formally, we proceed in three steps: 
\begin{enumerate*}[label=(\textit{\roman*})]
\item use a $k$-means algorithm on the \texttt{[CLS]} DistilBERT representations of MS MARCO queries to build 100 initial clusters\footnote{We consider a pre-trained DistilBERT~\cite{Sanh2019} that has not been fine-tuned.},
\item select the five clusters that maximize the sum of pairwise $\ell_2$ distances between their respective centroids, \item expand those five native clusters by joining nearest clusters, until we have groups of $\approx$25k queries (without overlap). 
\end{enumerate*}
This process differs from a classical k-mean algorithm as the final clusters do not form a partition of the entire set of queries ($|\bigcup_{i} C_i| \simeq 125k<<500k$), resulting in larger boundaries between clusters compared to the works from \cite{https://doi.org/10.48550/arxiv.2204.11447}.
Additionally, starting from $100$ clusters enables us to define more specific topics from the start (we also experimented with larger values for $k$, but didn't notice any improvements in the quality of the clustering). 
Finally, we split each cluster into \textbf{train} and \textbf{test} sets, allowing us to compare both in-domain and out-of-domain performance. Training is done on the train set of each $\overline{C_i}$, referred to as $\overline{C_i}^{t}$. It contains $\approx 100k$ queries, for which we sample 100 negatives using BM25, resulting in approximately 10M triplets in total. Models are evaluated on the test sets of the $C_i$ (referred to as $C_i^{e}$, containing $6200$ queries each). We provide a t-SNE~\cite{vanDerMaaten2008} visualization of the resulting clusters in Figure~\ref{fig:tsne}: we observe that clusters are clearly distinct from each other in the embedding space. 
We additionally provide examples from each cluster in Table~\ref{tab:query}: clusters correspond to different topics, so the distribution shift here is more semantic.

\paragraph{\bf{WH-Words Queries}} Besides query topics, we investigate queries styles and goals, through an analysis of question words~\cite{10.3115/1073012.1073082,https://doi.org/10.48550/arxiv.2204.12755}. In comparison to Natural Language Processing, question words in IR are a much stronger signal, as they define the query intent  recently studied in ~\cite{10.1145/3477495.3531926}, such as:  \textit{instruction}, \textit{reason}, \textit{evidence-based}, \textit{comparison,} \textit{experience} or \textit{debate}. In order to evaluate models on such shifts, we manually build three clusters ($wha$, $how$, $who$, also referred to as $(W_i)_{i=0,1,2}$), for queries respectively related to definitions (``what", ``definition"), instructions (``how") and finally more general questions linked to persons, locations or context (``who", ``when", ``where", ``which"). This shift is entirely rule-based, as we separate queries on the above lists of fixed terms. Similarly, we split each cluster into \textbf{train}/\textbf{test} splits, and we train models on the training sets of the complements (containing $10M$ triplets in total). We evaluate models on the test $W_i$, each containing 6500 queries. 

\paragraph{\bf{Short and Long Queries}} Query length is known to greatly impact retrieval -- for instance, aggregating information from long queries has been deemed difficult for both traditional and neural IR methods~\cite{beir_2021}. We thus define the last shift on this attribute. To analyse this effect, we split the train set into groups of short and long queries, from the median query length at the word level ($6$ for MS MARCO). 
\textbf{Train}/\textbf{test} sets contain respectively $10M$ training triplets and $3500$ queries for evaluation.

\begin{figure}[t]
    \centering
	\includegraphics[width=0.6\columnwidth]{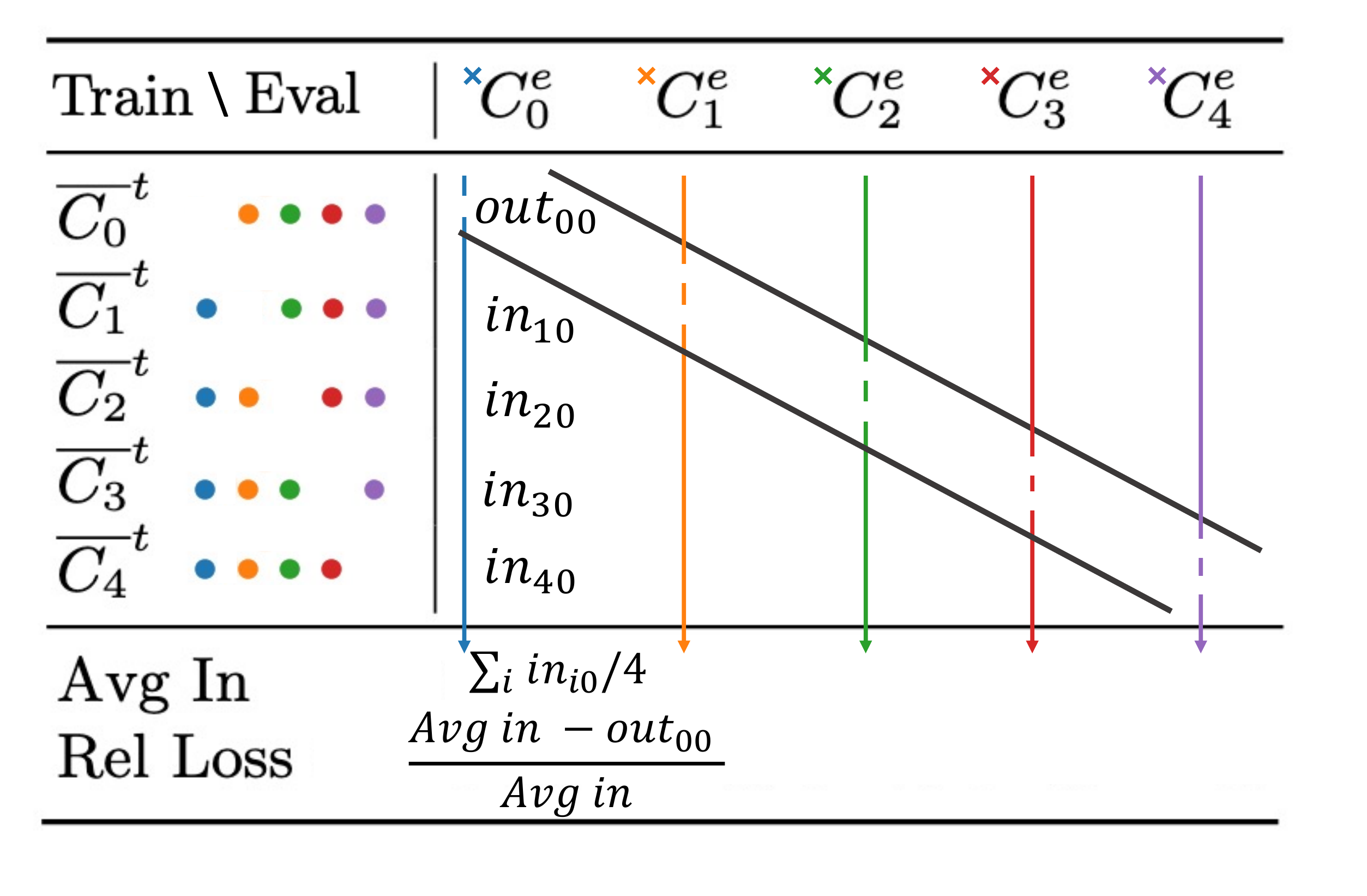}
	\caption{Zero-shot evaluation procedure. Lines correspond to models trained on different clusters, and columns to evaluation sets. For $C_0$, we define \textit{Avg In} as the mean test performance on $C_0^{e}$ for models trained on $\overline{C_1}^{t}, \overline{C_2}^{t}, \overline{C_3}^{t}$ and $\overline{C_4}^{t}$, and \textit{Rel Loss} as the relative difference between the \textit{Avg In} and the zero-shot performance (i.e. trained on $\overline{C_0}^{t}$ and evaluated on $C_0^{e}$).}
	\label{fig:l1o}
\end{figure}

\subsection{Evaluation procedure}

We perform leave-one-out on all the shifts in order to evaluate the in-domain and zero-shot effectiveness of various models. For instance, let's consider the topic clusters. For $i \in \llbracket 0,4\rrbracket$, we independently train a model $\mathcal{M}_i$ on $\overline{C_i}^{t}$, and evaluate it on the test set of each cluster. It, therefore, creates a zero-shot experiment on $C_i^{e}$, as its distribution was out-of-training. We also have access to in-domain evaluation on the $C_j^{e}$ for $j\ne i$. Similarly with wh-words, we train on $\overline{wha}^{t}$, $\overline{how}^{t}$ and $\overline{who}^{t}$, and test each time on respective test sets. Note that, in addition to the leave-one-out rotation on clusters, we also rely on train/test splits inside each cluster, in order to have access to both in-domain and out-of-domain performances. We, therefore, report \textit{Avg In} as the average performance measure, when the distribution of the evaluated cluster is seen at training time, and \textit{Rel Loss} as the relative loss between the above average measure, and the zero-shot performance, i.e. when the evaluated cluster is out of the training distribution. Figure~\ref{fig:l1o} illustrates the overall evaluation procedure. Note that, as some clusters may intrinsically contain harder queries, what we are really interested in here is the comparison of the performance inside a column (i.e., on the same evaluation set, with different training sets). In general, the lowest performance is achieved on the diagonal (as it corresponds to zero-shot evaluation).

\section{Experimental Setup}\label{sec:exp}


We compare three first-stage ranking models: \begin{enumerate*}[label=(\textit{\roman*})]
    \item a standard dense \texttt{bi-encoder}~\cite{karpukhin2020dense} which represents queries and documents in a dense low dimensional space by the means of the \texttt{[CLS]} embedding,
    \item the late-interaction \texttt{ColBERT}~\cite{colbert},
    \item and the sparse bi-encoder \texttt{SPLADE-max}~\cite{formal2021splade} which represents queries and documents as sparse high-dimensional bag-of-words vectors.
\end{enumerate*} Those models represent the three main families of representation-based models in IR, and we are thus interested in their behavior with respect to zero-shot generalization.
Finally, we also include BM25  as a reference point, as well as a \texttt{monoBERT} cross-encoder~\cite{passage_ranking} re-ranking BM25 top-$1000$ documents.
All models rely on a pre-trained DistilBERT~\cite{Sanh2019} backbone model, and are fine-tuned on a particular query subset. We limit ourselves to a standard training procedure, relying on contrastive learning and BM25/in-batch negatives~\cite{karpukhin2020dense}, without further improvements such as distillation~\cite{https://doi.org/10.48550/arxiv.2010.02666}, hard negative mining~\cite{xiong2021approximate} or middle-training~\cite{gao-callan-2022-unsupervised}, as those techniques are more general, and 
may be applied to any baseline model. 
For evaluation, we report MRR@10 -- the official MS MARCO performance measure. We also report the Atomized Search Length\footnote{We used a 100 bounded ASL.} (ASL)~\cite{https://doi.org/10.48550/arxiv.2201.01745}, as it complements MRR@10 by not focusing on top of the ranking. 
In a nutshell, it is defined, for a given query, as the average number of irrelevant documents ranked before a relevant one. 
The dense and sparse bi-encoders are fine-tuned on $4$ V100 GPUs, for $100k$ iterations, with a batch size of 128, using the MS MARCO triplets from our shifts. ColBERT and monoBERT are both trained for $150k$ iterations with a default batch size of $32$, on 2 V100 GPUs. 
For the bi-encoders, best checkpoints are selected using an approximated early stopping ~\cite{Hofstaetter2021_tasb_dense_retrieval} relying on a validation set composed of 1600 queries, which is \emph{not} subject to the shifts. From our observations, the best checkpoints would generally correspond to 
$40k$ iterations, resulting in about $5M$ training triplets (using $bs=128$). ColBERT and monoBERT, on the other hand, do not rely on early stopping, but overall see the same number of training samples ($5M$, in $150k$ iterations with $bs=32$). Thus, both training procedures are very similar. For all other parameters, we adopt the default values reported in the original papers.


\section{Results and Analysis}\label{sec:res}

In this section, we first compare the drops in performance of different architectures when subject to the shifts. We then link those results with two indicators we define in section~\ref{sec:similarity}, which can be used to analyse the behavior of different models regarding their zero-shot capability.

\subsection{Performance Evaluation on Distribution Shifts} 


\begin{table}[t]
\begin{center}
    \caption{Comparison of the average performance and relative loss (MRR@10) from seen to unseen clusters. In bold are the best on each cluster (in terms of performance and loss). All losses between the \textit{Avg In} and \textit{Out}, for each model, are statistically significant with $p$-value$<0.05$ for paired $t$-test.}
    \label{tab:delta}
    \begin{tabular}{l  l  c c c c c}
        \toprule
        \textbf{Models}  & & $C_0^e$ & $C_1^e$ & $C_2^e$ & $C_3^e$ & $C_4^e$  \\ \midrule
        BM25 & & 19.2  & 25.9 &	16.4 & 18.1  & 17.5  \\ \midrule
        bi-encoder & Avg In& 33.2 &	37.2&	28.5 &	21.5 &	21.4 \\ 
         & Out & 30.4 & 30.5 & 25.9 & 19.0 & 19.6 \\
        & Rel Loss & 8.3\%  &18.0\% &	9.0\% &11.5\%  &	8.6\%  \\  \midrule
        SPLADE & Avg In & 36.8 & 38.7 & 31.2 & 26.2 & 25.0 \\
        & Out & 34.5 & 34.0 & 30.2 & 24.5 & 24.7 \\
         & Rel Loss & 6.3\% & 12.2\% & \textbf{3.2\%} & 6.4\% & \textbf{1.4\%}  \\ \midrule
        ColBERT & Avg In& \textbf{39.7} & 42.3 &\textbf{34.6}&\textbf{28.8}&\textbf{27.7} \\ 
        & Out & \textbf{38.6} & \textbf{38.7} & \textbf{33.4} & \textbf{27.7} & \textbf{27.1} \\
        & Rel Loss & 2.7\% & \textbf{8.5}\% & 3.4\% & \textbf{3.7}\% & 2.2\%  \\ \midrule
        monoBERT & Avg In & 39.4 & \textbf{42.7} & 33.4 & 27.1 & 26.3 \\
        & Out & \textbf{38.6} & 38.4 & 31.8 & 25.9 & 25.7 \\
         & Rel Loss & \textbf{2.1\%} & 10.2\% & 4.8\% & 4.5\% & 2.4\%  \\ 
        \bottomrule
    \end{tabular}   
\end{center}
\end{table}


\paragraph{\bf{Query Topics}} Table~\ref{tab:delta} reports the performance of models on both in-domain (\textit{Avg In}) and out-of-domain (\textit{Out}), as well as the relative (\textit{Rel Loss}) due to the shift, on each topic cluster. We first notice that drops in performance can be significant (up to $18$\% for the dense bi-encoder). Overall, we observe that dense models are the most impacted by the shifts across clusters, followed by SPLADE, and finally ColBERT and monoBERT. 
Interaction approaches also demonstrate both better performance on in-domain and out-of-domain -- compared to representation-based models. All in all, the results are in line with the ones reported in~\cite{beir_2021}. However, contrary to the results reported in \cite{https://doi.org/10.48550/arxiv.2204.11447}, SPLADE seems here to be substantially more robust than dense bi-encoders, and this on every cluster. Note that, the average query/document sizes for SPLADE (which correspond to the number of non-zero dimensions in the sparse representations) are respectively 24 and 130, which is already way below the dense representations (\texttt{[CLS]} vector of size $768$) -- despite its better performance overall. 
Moreover, as opposed to the BEIR benchmark, no model here performs lower than BM25 under zero-shot evaluation. This phenomenon is likely due to the fact that we ``stay'' within MS MARCO, tackling the same retrieval task. 
BM25 performance also indicates to some extent the difficulty of each cluster -- some of them (e.g. $C_1^{e}$) supposedly relying more on word matching than others. Finally, note that all the drops (between in- and out-of-domain) are statistically significant, with $p$-values$<0.05$.


\begin{table}[t]
\begin{center}
    \caption{Comparison of the avg. MRR@10 and relative losses in zero-shot for wh-words and query length. All losses have $p$-value$<0.05$ for paired $t$-test.}
    \label{tab:wh}
    \begin{tabular}{l  l  c c c |c c}
        \toprule
        \textbf{Models} & & $wha^e$ & $how^e$ & $who^e$ & $short^e$ & $long^e$  \\ \midrule
        BM25 & & 18.2 & 14.7 & 22.3 & 19.0 & 18.5 \\ \midrule
        bi-encoder & Avg In & 27.8  & 26.0 & 33.1 & 34.0 &27.1\\ 
         & Out & 23.4 & 19.6 & 27.9 & 29.8 & 25.2 \\
        & Rel Loss & 15.8\% &24.8\%   &15.8\% & 12.5\% & 7.0\% \\   \midrule
        SPLADE & Avg In & 30.3&28.9&37.7 & 34.9 &30.3 \\
        & Out & 28.6 & 21.2 & 32.5 & 33.5 & 27.1 \\
         & Rel Loss &  5.5\% & 26.8\% & 13.7\% & \textbf{3.9\%} & 10.4\% \\ \midrule
        ColBERT & Avg In& 33.5 & \textbf{31.7} & 40.0 & \textbf{38.4} & 32.5 \\ 
        & Out & \textbf{31.8} & \textbf{27.3} & 36.6 & \textbf{36.4} & \textbf{31.6} \\
        & Rel Loss & \textbf{5.2\%} & 14.0\% & 8.6\% & 5.1\% & \textbf{2.7\%} \\ \midrule
        monoBERT & Avg In& \textbf{33.9}&30.5&\textbf{40.1} & 37.7 &  \textbf{32.6}  \\ 
        & Out & 31.1 & 26.4 & \textbf{37.1} & 32.3 & 28.9 \\
        & Rel Loss & 8.3\% & \textbf{13.5\%} & \textbf{7.5\%} & 14.3\% & 11.3\% \\
        \bottomrule
    \end{tabular}   
\end{center}
\end{table}


\paragraph{\bf{WH-Words Queries}}
Table~\ref{tab:wh} shows the results on wh-words and query length shifts. On the former (left columns), drops in effectiveness are on average much more important compared to topic clusters -- with up to 26.8\% on the $how^e$ cluster for SPLADE. However, the overall comparison between architectures remains the same. We also notice that models with higher \textit{Avg In} are also the ones with the lowest relative losses -- ruling out overfitting as the cause of better in-domain performance. Alexander et al.~\cite{2022arXiv220500926A} conduct a study on query taxonomies. In particular, they refer to $how$ queries as having an \textit{instrumental} intent ("how", "how to", "how do"), in contrast to usual queries which have \textit{factual intents} (e.g. "who", "when", "where", "which", "what", "definition"). We notice in our results the same pattern on those queries, with both higher drops, as well as lower BM25 results. Those queries tend to rely less on word matching, and more on a general understanding of the given situation -- it is thus harder for models to accurately perform on such a cluster in zero-shot.

\paragraph{\bf{Length}}
Concerning the length-based shift (Table~\ref{tab:wh}, right columns), we first see that short queries are easier than longer ones: a model trained on long queries will be better on the $short$ than on the $long$ evaluation set. 
Interestingly, we notice that short and long queries are somewhat complementary in a training set -- a model cannot be trained with long queries only. Besides, we also observe that models trained on long queries tend to have a higher recall (not reported), while models trained on short queries a higher precision. 



\subsection{Train/Test Distribution Similarity}
\label{sec:similarity}

From our previous experiments, it is difficult to estimate \textit{a priori} the strength of a shift on the downstream performance. We thus consider in the following two measures of similarities between train and test queries, that partially correlate with the zero-shot performance drop -- and so the strength of the shift -- and that can easily be interpreted: \begin{enumerate*}[label=(\textit{\roman*})] 
\item a term-based similarity, which measures the vocabulary overlap between out-of-domain and in-domain query sets,
\item a model-based similarity, which takes advantage of internal representations of trained models.
\end{enumerate*} 

\paragraph{\bf{Jaccard Similarity}} The weighted Jaccard~\cite{w_jaccard} can be used to measure the vocabulary overlap between two sets of documents or queries~\cite{10.1145/3488560.3498406,beir_2021}. More formally, given a source and a target collection, it is defined as:
$$
J(S, T) = \frac{\sum_V\min(S_k, T_k)}{\sum_V\max(S_k, T_k)}
$$
where $S_k$ and $T_k$ are the normalized frequencies of word $k$ in source and target datasets respectively, and $V$ is the union vocabulary. In our case, we aim to measure the similarity $J(C_i, \overline{C_i})$ between queries from cluster $C_i$ (out-of-domain) and its complement $\overline{C_i}$ (in-domain)\footnote{Formally, we should compute $J(C_i^e, \overline{C_i}^t)$, but terms statistics at the cluster level for train and eval sets are very similar.}. 
Intuitively, we are trying to quantity to which extent a given out-of-domain query set is \emph{far} for the training set, based on term statistics. This metric is computed at the cluster level, such that we can associate it with the overall out-of-domain loss.
\begin{figure}[t]
    \centering
	\includegraphics[width=0.6\columnwidth]{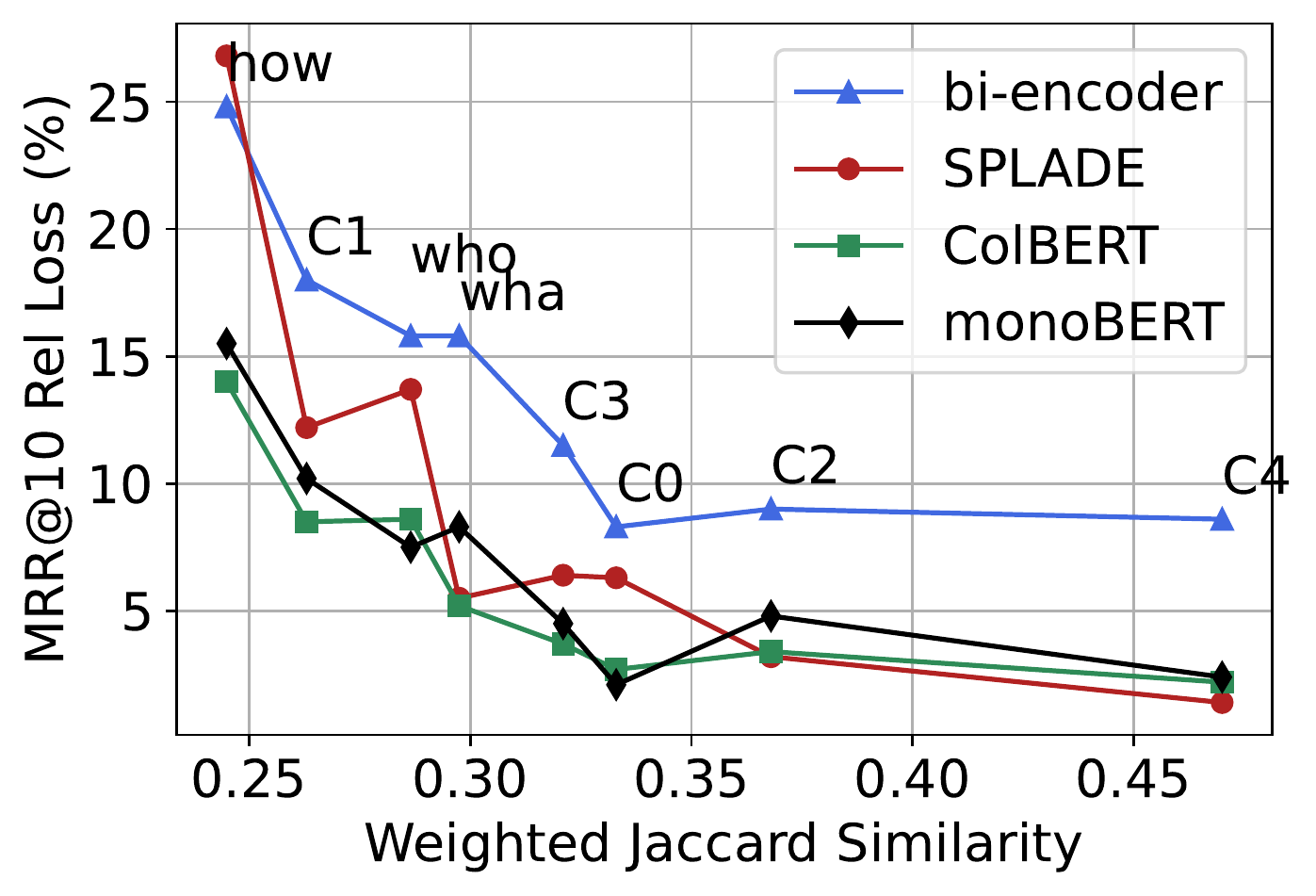}
	\caption{Relative loss on zero-shot clusters with respect to weighted Jaccard similarities between train and test query distributions. Leftmost part: low similarities and high losses. Rightmost part: high similarities and small losses.}
	\label{fig:jacc}
\end{figure}

We plot in Figure~\ref{fig:jacc} the relative loss on each of the out-of-domain cluster, with respect to the weighted Jaccard $J(C_i, \overline{C_i})$ and $J(W_i, \overline{W_i})$. From the figure, we observe that, as the Jaccard similarity increases -- and so, the terms overlap -- the relative loss in performance diminishes. 
Such a measure is thus indicative when it comes to predicting generalization capabilities of neural models. 
An important point is that the behavior is common for both Query Topics and WH-Words Queries, suggesting that this pattern may be independent of the nature of the shifts.
With the hypothesis that all models still partly rely on term matching, 
it also shows that they tend to have issues when learning the general pattern of word matching, independently from the terms themselves. 
This extends the observation made for dense models in~\cite{https://doi.org/10.48550/arxiv.2204.12755}, and supports~\cite{formal2021match} about generalization to unseen words.
We further notice that, for the clusters with the highest Jaccard ($C_2^e$ and $C_4^e$), the relative loss is lower for SPLADE, compared to ColBERT. We hypothesize that, as sparse models rely more on lexical components (through the BoW representation), they are better able to transfer when the query vocabulary distributions are closer.
Finally, the same relation can be observed for ASL (instead MRR@10) -- although not reported here -- indicating the quality of the indicator for recall-oriented metrics.

\paragraph{\bf{Model-based Similarity}} 
To complement the above lexical indicator, we additionally introduce a semantic measure of similarity, which relies on the internal representations of a trained bi-encoder. 
Intuitively, for a model trained on a given set of queries, we compute the distances between those training queries and a targeted test query.
More formally, we compute the mean retrieval score of a dense bi-encoder $\mathcal{M}_i$ between a test query $q^e \in C_i^e$ and the training queries from $\overline{C_i}^t$, as follows:
\begin{equation}\label{eq:1}
R(q^e, \overline{C_i}^t)=\frac{1}{|\overline{C_i}^t|}\sum_{q^t \in \overline{C_i}^t} s_{\mathcal{M}_i}(q^e, q^t)
\end{equation}
where $s_{\mathcal{M}_i}$ is the output score of the dense bi-encoder $\mathcal{M}_i$ trained on $\overline{C_i}^t$ (in our case, $s_{\mathcal{M}_i}$ is a dot product between query embeddings). We use \emph{trained} dense bi-encoders as the baseline for the similarity computation, as their symmetric nature makes it more natural to compute similarities between queries. This representation-based similarity thus enables to quantify how \textit{far} is a test query
 from the training set. Contrary to the Jaccard similarity, this indicator is, however, defined \textit{at the query level} -- such that it is possible to link it to the loss associated with each query. Moving away from cluster-based to query-level indicators is interesting from a practical standpoint. Given a query from a new domain, we would like to be able to infer model performance. We note the interesting parallel that can be made with the vast literature around Query Performance Prediction, where the goal is to estimate the performance of an IR system without relevance judgements~\cite{10.1145/1835449.1835683}. 

\begin{figure}[t]
    \begin{minipage}{0.49\textwidth}
        \centering
        \includegraphics[width=1\linewidth]{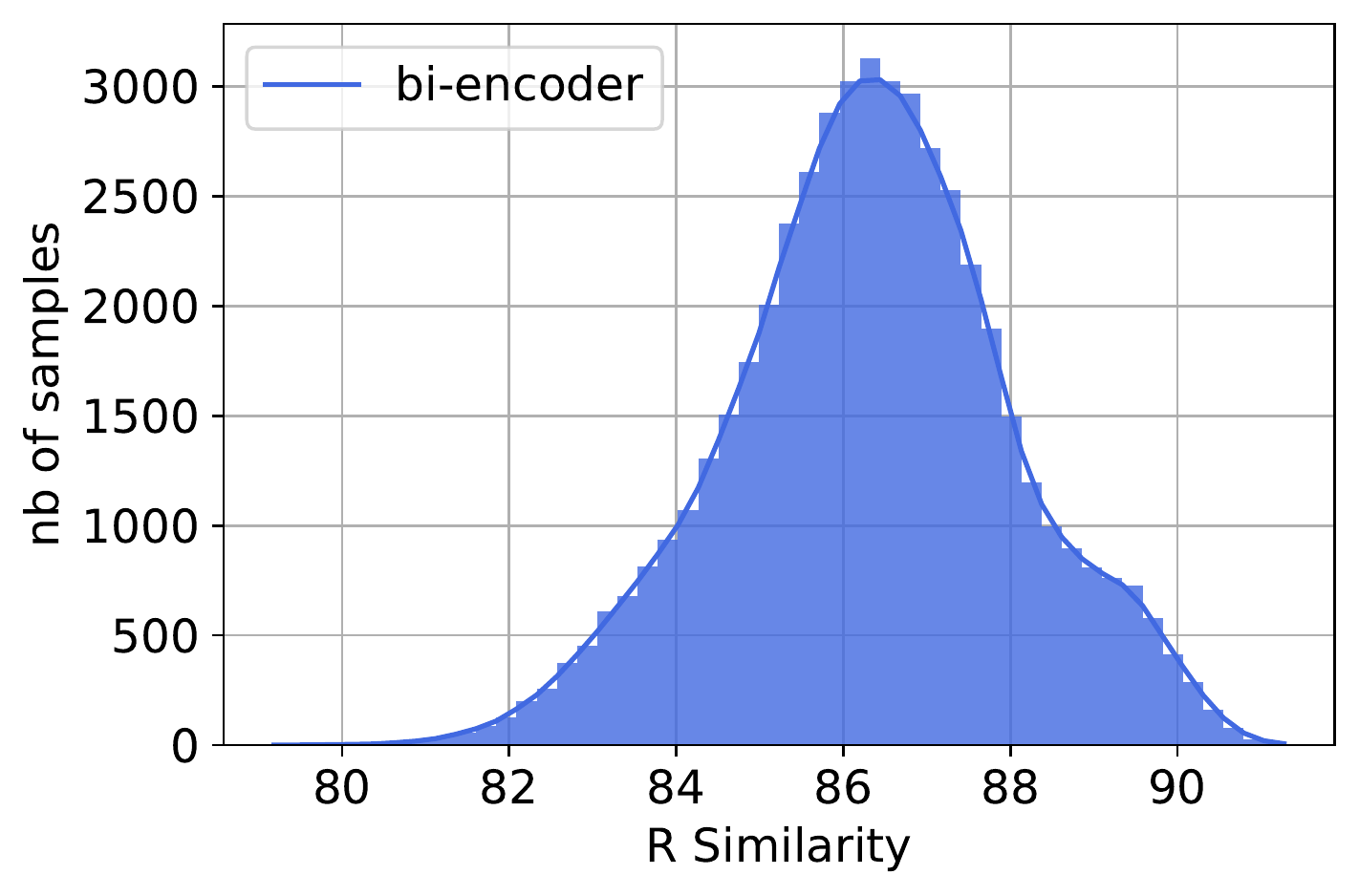}
        \caption{Distributions of query scores. $R$ similarity gives an estimation of how close each query is to the training set.}
        \label{fig:distribution}
    \end{minipage}\hfill
    \begin{minipage}{0.49\textwidth}
        \centering
        \includegraphics[width=1\linewidth]{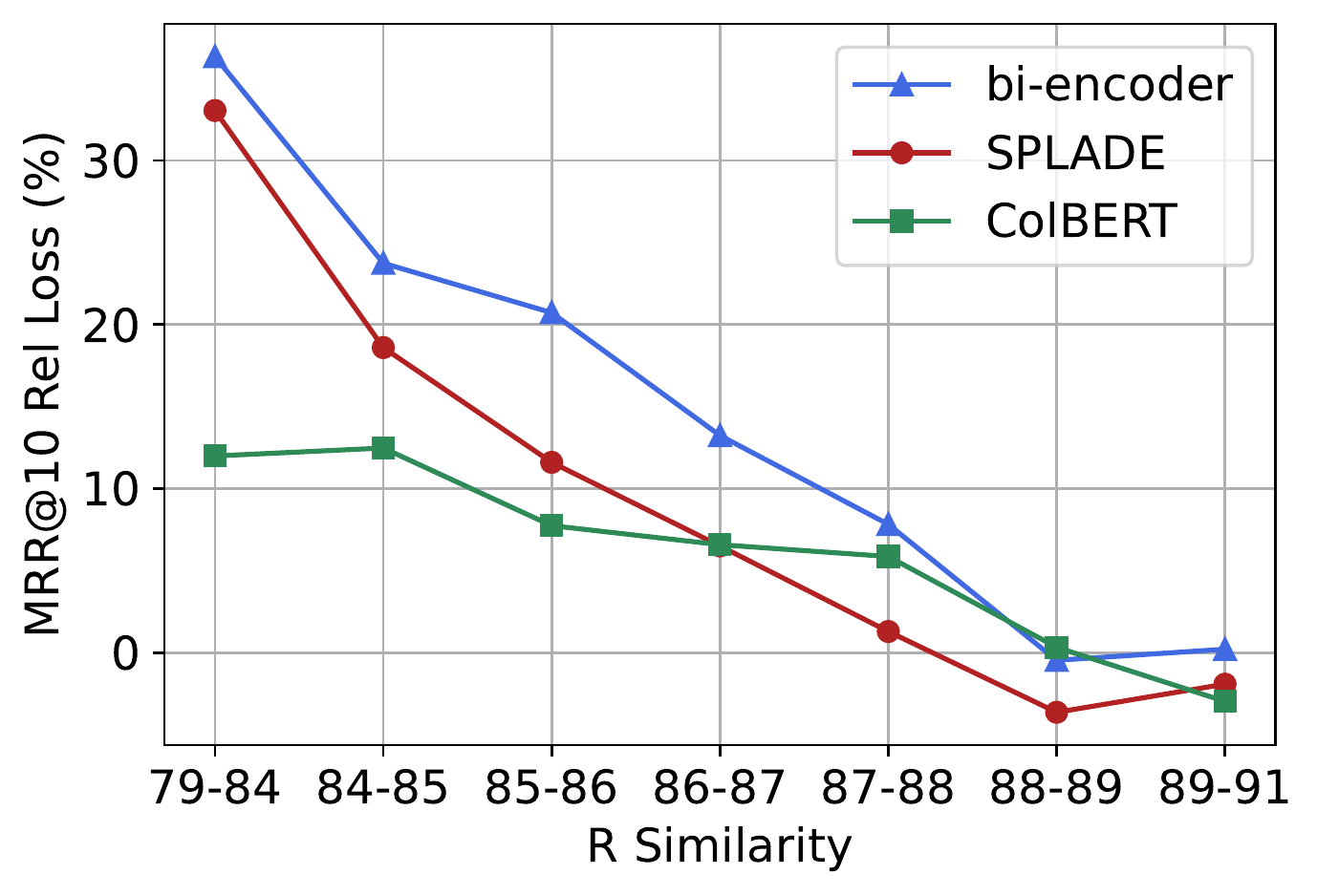}
        \caption{Relative MRR@10 loss on zero-shot queries with respect to different intervals of $R$ similarities.}
        \label{fig:meanqpp}
    \end{minipage}\hfill
\end{figure}
\begin{figure}[t]
   \begin{minipage}{0.49\textwidth}
     \centering
     \includegraphics[width=1\linewidth]{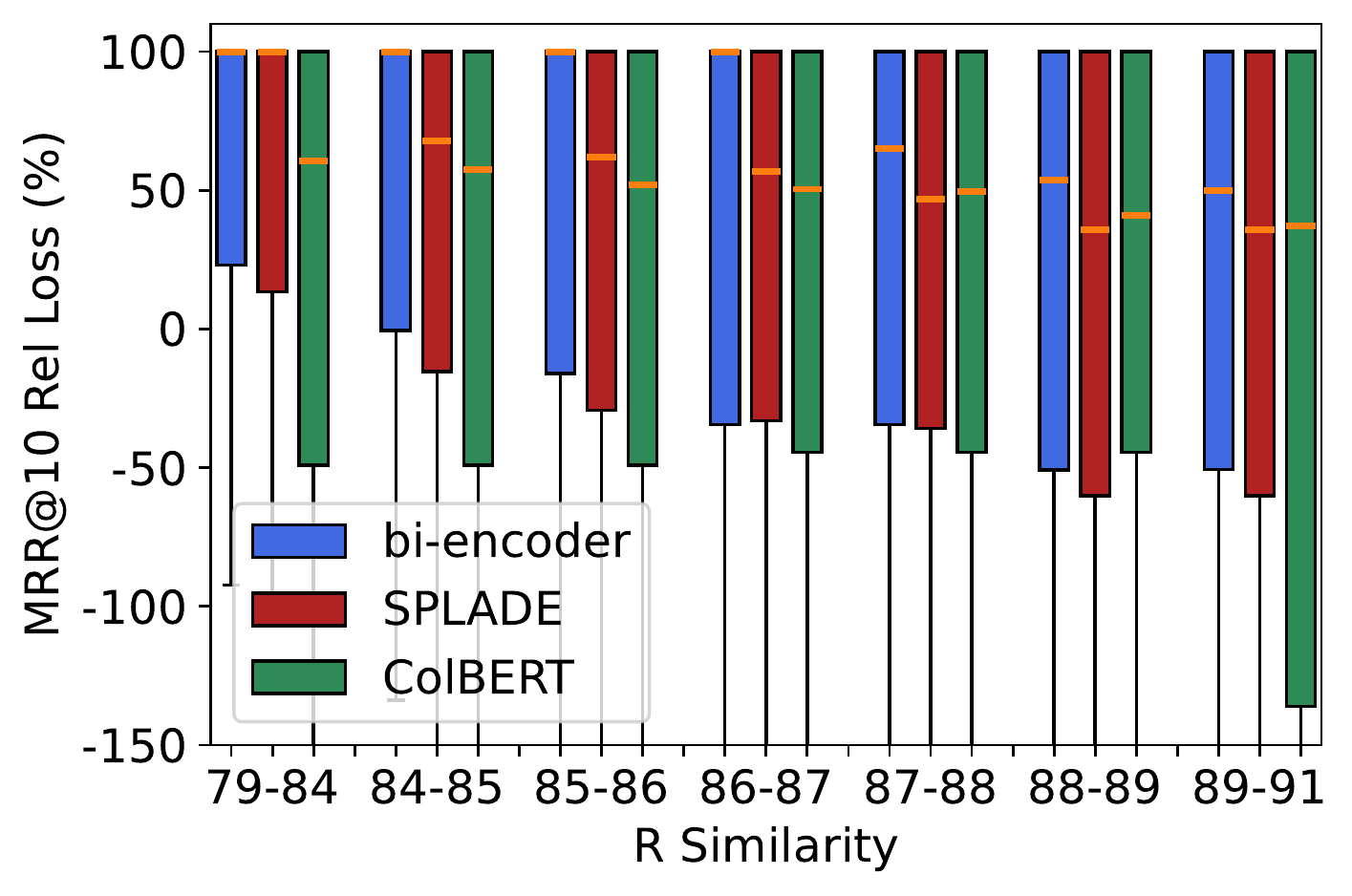}
   \end{minipage}\hfill
   \begin{minipage}{0.49\textwidth}
     \centering
     \includegraphics[width=1\linewidth]{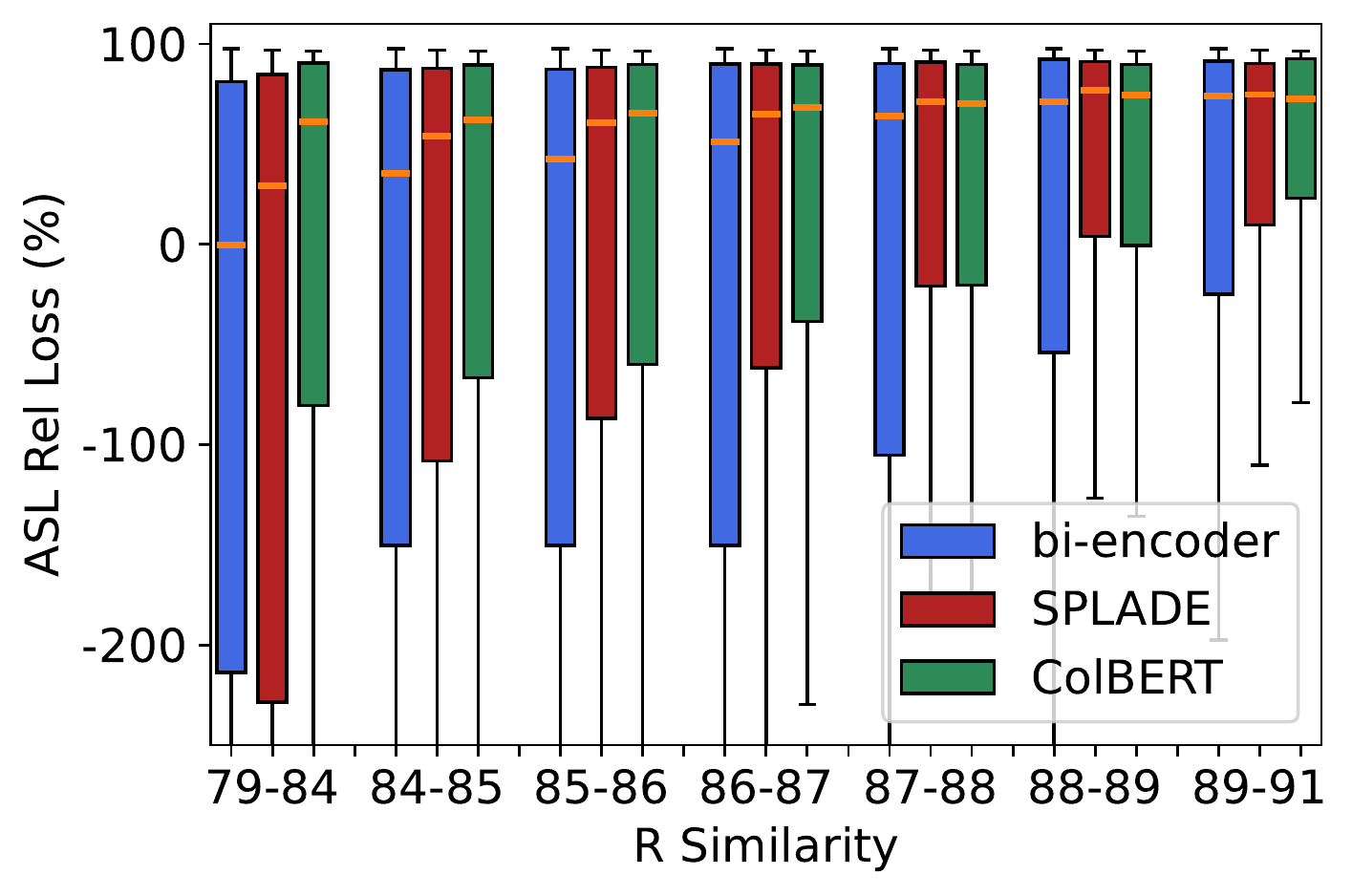}
   \end{minipage}
    \caption{Boxplots of the MRR@10 and ASL relative loss with respect to $R$ similarity (Median indicated in orange). Left: MRR@10 ($\uparrow$). Right: For ASL ($\downarrow$), the lower the better, so the higher the relative losses, the better. Both metrics indicate better performances on queries close to the training distribution (rightmost parts of both graphs).}
    \label{fig:boxplotqpp}
\end{figure}

We show on Figure~\ref{fig:distribution} the distributions of the above similarity $R(q^e, \overline{C_i}^t)$ and $R(q^e, \overline{W_i}^t)$ for all $\{(q^e, \overline{C_i}^t) | i \in \llbracket 0,4\rrbracket, q^e \in C_i^e\}$ and $\{(q^e, \overline{W_i}^t) | i \in \llbracket 0,2\rrbracket, q^e \in W_i^e \}$, for the dense bi-encoder.
Then, in Figure~\ref{fig:meanqpp}, we plot the average relative loss in terms of MRR@10 with respect to different intervals of $R$ similarities -- corresponding to zero-shot queries from topics and wh-words clusters. Overall, those intervals represent the spectrum of queries that are the further away from the train set (leftmost part, \textit{low similarity}) to the closest ones (rightmost part, \textit{high similarity}). For the three architectures, the farther the out-of-domain test query is, the highest the loss. When comparing models, ColBERT seems to generalize better on the most distant queries compared to SPLADE and the dense bi-encoder.
Results are aligned with the ones observed for the Jaccard -- but for a semantic notion of distance. It is also interesting to see that the trend holds for SPLADE and ColBERT, given that the similarity is entirely based on the dense bi-encoder representations. We can thus infer that the knowledge accumulated by dense bi-encoders at training time could be used as a potential signal to predict performance, without additional supervision.

In addition, in Figure~\ref{fig:boxplotqpp}, we represent the same intervals with boxplots, to analyse the variance of the MRR@10 and ASL (as those metrics are initially defined at the query level). 
Looking at the variance of the MRR@10, we see that it decreases as we go further away from the training set, implying that the losses on the top of the ranking will be high for distant queries.
On the other hand, with ASL, the variance behaves differently: we have high confidence in the closest queries, on which the ASL will improve (decrease), while performance on distant queries is uncertain. 
ASL being a more recall-oriented metric, both results are thus complementary and give an overview of the behavior at both the very top of the ranking with MRR@10, and at deeper ranks with ASL.

\section{Conclusion}\label{sec:conclu}

In this work, we focus on zero-shot evaluation of neural IR models. We propose to benchmark neural retrievers based on PLM against three controlled distribution shifts, created by partitioning MS MARCO training queries -- based on different characteristics (semantic, intent, and length). Overall, we observe that interaction approaches are more robust than representation-based models, and this across all shifts. We further link the observed drops in performance to two indicators which verify that, the further away the test set is from the train one, the worse the drop in performance.
Our analysis seems to suggest that a model-based similarity could possibly be used as \emph{unsupervised} predictors of performance. We believe it opens the path to future research directions that need to be investigated in, and foster open questions for the IR community on how to measure model robustness. 
Furthermore, the effect of techniques such as distillation or pre-training remains yet to be analysed, as they could possibly correct model biases and lead to better generalization. 




\section*{Acknowledgements}

We first would like to thank Carlos Lassance, Hervé Déjean and Jean-Michel Renders for the valuable discussions and feedback on the paper. We also would like to thank Guglielmo Faggioli, Stefano Marchesin, Nicola Ferro and Benjamin Piwowarski a lot, for the knowledge they provided us on Query Performance Prediction.

%
%
%
%
\bibliographystyle{splncs04}
\bibliography{my-bib}






\end{document}